\documentclass[journal]{IEEEtran}
\usepackage[caption=false,font=footnotesize,labelfont=sf,textfont=sf]{subfig}
\usepackage{bbding}
\usepackage{multirow}

\usepackage{setspace}
\usepackage{cite,graphicx,amsmath,amssymb}
\usepackage{fancyhdr}
\usepackage{mdwmath}
\usepackage{mdwtab}
\usepackage{balance}
\usepackage{xcolor}
\usepackage{bm}
\usepackage{amsthm}
\usepackage{threeparttable}
\usepackage{algorithm}
\usepackage{algorithmic}
\usepackage{multirow}
\usepackage{flafter}
\usepackage{makecell}
\usepackage{diagbox}
\usepackage{caption}

\usepackage{hyperref}

\providecommand{\url}[1]{#1}
\allowdisplaybreaks
\begin{document}

\title{Artificial Intelligence Enabled NOMA Towards Next Generation Multiple Access}

\author{
Xiaoxia~Xu,
Yuanwei~Liu,
Xidong~Mu, 
Qimei~Chen,
Hao~Jiang,
and Zhiguo~Ding

\vspace{-1em}

\thanks{X. Xu, Q. Chen, and H. Jiang are with the School of Electronic Information, Wuhan University, Wuhan, 430072, China (e-mail: \{xiaoxiaxu, chenqimei, jh\}@whu.edu.cn).}
\thanks{Y. Liu and X. Mu are with the School of Electronic Engineering and Computer Science, Queen Mary University of
London, London E1 4NS, U.K. (email: \{yuanwei.liu, xidong.mu\}@qmul.ac.uk).}
\thanks{Z. Ding is with the School of Electrical and Electronic Engineering, The University of Manchester, Manchester M13 9PL, U.K. (email: zhiguo.ding@manchester.ac.uk).}
}

\maketitle

\begin{abstract}
This article focuses on the application of artificial intelligence (AI) in non-orthogonal multiple-access (NOMA), which aims to achieve automated, adaptive, and high-efficiency multi-user communications towards next generation multiple access (NGMA). First, the limitations of current scenario-specific multiple-antenna NOMA schemes are discussed, and the importance of AI for NGMA is highlighted.  Then, to achieve the vision of NGMA, a novel cluster-free NOMA framework is proposed for providing \emph{scenario-adaptive} NOMA communications, and several promising machine learning solutions are identified. To elaborate further, novel centralized and distributed machine learning paradigms are conceived for efficiently employing the proposed cluster-free NOMA framework in single-cell and multi-cell networks, where numerical results are provided to demonstrate the effectiveness. Furthermore, the interplays between the proposed cluster-free NOMA and emerging wireless techniques are presented. Finally, several open research issues of AI enabled NGMA are discussed.
\end{abstract}

\section{Introduction}
Wireless systems are witnessing an explosive proliferation of cellular devices and emerging wireless applications.
To support high data speed and massive connectivity over limited spectrum resources, it is crucial to investigate advanced multiple access technologies for next generation wireless systems, namely next generation multiple access (NGMA)\cite{NGMA2022}.
As a promising multiple access technique, non-orthogonal multiple access (NOMA) has been considered as an indispensable component for NGMA.
In particular, the integration of NOMA and multiple-antenna technology has attracted significant attentions and demonstrated tremendous potentials in connectivity and capacity improvements\cite{BBNOMA_Chen,MIMONOMA_2016}.
However, to deal with the highly diversified communication scenarios and overwhelming system complexity of next generation wireless systems, it is imperative to explore adaptive, automated, and intelligent communication designs for NGMA.

\begin{figure*}[!h]
  \vspace{-1.5em}
  \centering
  \includegraphics[width=0.88\textwidth]{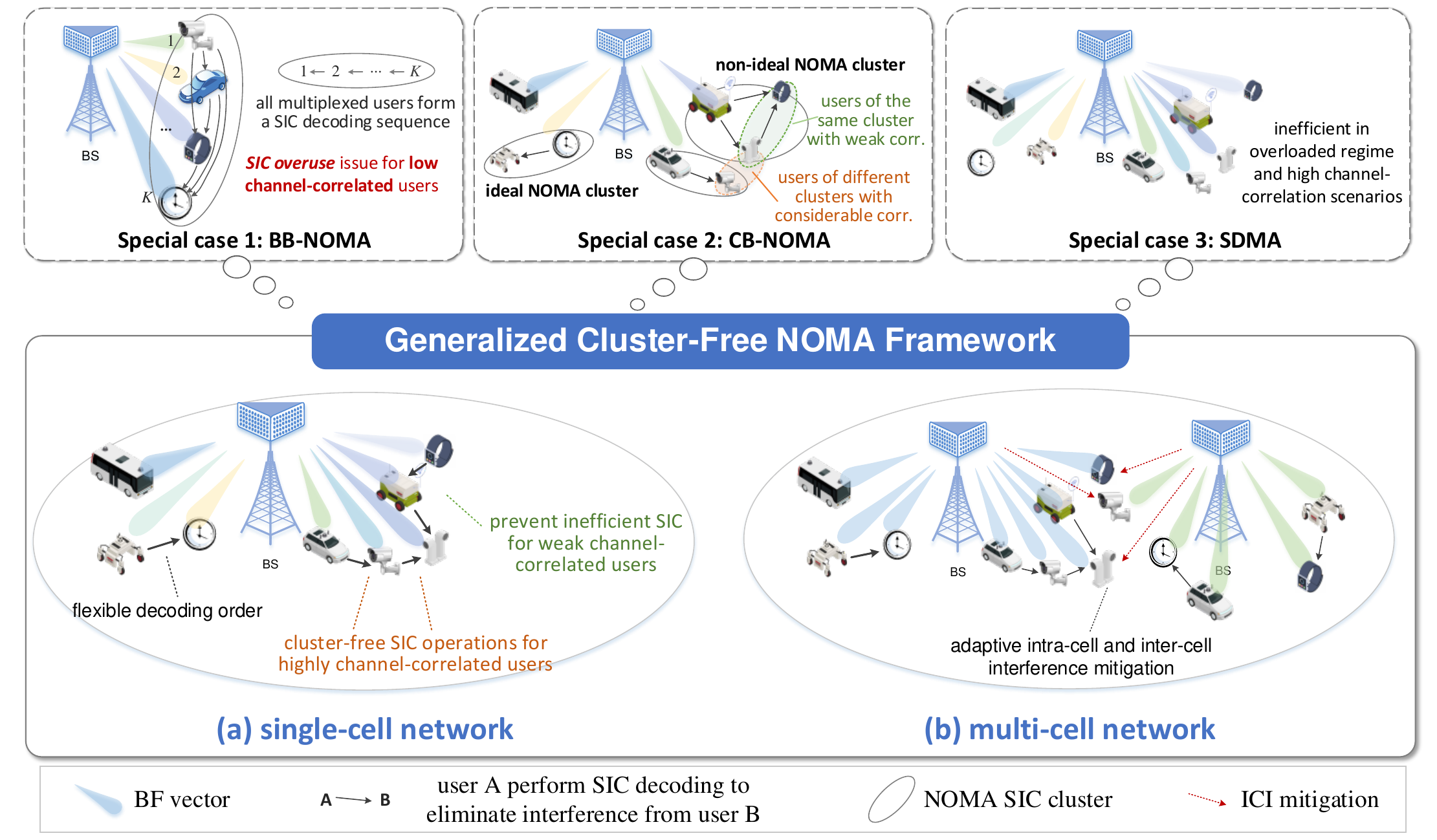}\\
  \caption{Illustration of the proposed generalized cluster-free NOMA framework in single-cell and multi-cell networks.}\label{fig_sys_model}
\end{figure*}

\subsection{From NOMA to NGMA}

To drive the evolution of NOMA towards NGMA, we start by rethinking the limitations of existing multiple-antenna NOMA approaches.
Existing multiple-antenna NOMA approaches can be mainly categorized into beamformer-based NOMA (BB-NOMA) and cluster-based NOMA (CB-NOMA). 
The main difference between these approaches lies in the successive interference cancellation (SIC) and beamforming designs \cite{NGMA2022}.
To be more specific, BB-NOMA allocates dedicated beamforming vector to each user, 
and then performs SIC in a certain decoding order to suppress the residual spatial interference \cite{BBNOMA_Chen}.
Nonetheless, since each high-order user needs to decode the signals of all low-order users, BB-NOMA suffers from high SIC decoding complexity in the overloaded regime.
Moreover, it leads to the problem of \textit{SIC overuse} when users have low channel correlations \cite{NGMA2022}.
That is, for two users with low-correlated or orthogonal channels, the interference could be straightforwardly suppressed via spatial beamforming, 
while employing SIC would impair the system performance due to the requisite SIC decoding condition.
Compared to BB-NOMA, CB-NOMA separates users into multiple clusters according to their spatial channel correlations \cite{MIMONOMA_2016}, 
where the number of clusters is usually the same as the number of radio frequency (RF) chains.
By serving each cluster with the same beamforming vector and sequentially carrying out SIC within each cluster, 
CB-NOMA can reduce the system complexity and alleviate the \textit{SIC overuse} problem.
However, it relies on the assumption that users can be partitioned into multiple ideal clusters, such that users of the same cluster have high channel correlations, 
whilst users from different clusters are low channel-correlated.
But this ideal assumption may not always be achievable due to the channel randomness.

Intuitively, both SIC strategies of BB-NOMA and CB-NOMA are cluster-dependent, 
where users should be assigned into one or multiple clusters and then sequentially perform SIC within the clusters.
Nonetheless, as discussed above, both BB-NOMA and CB-NOMA are effective for specific scenarios, 
which cannot adaptively cope with diverse heterogeneous communication scenarios faced by next generation wireless systems. 
Against this background, we propose a novel downlink generalized cluster-free NOMA framework in this article. 
By releasing the cluster restrictions, ultra-flexible SIC can be achieved to efficiently mitigate interference and improve system performances. 
The proposed framework can unify existing spatial-division multiple access (SDMA), BB-NOMA, and CB-NOMA schemes, 
thus reaping their gains as well as overcoming their drawbacks in various scenarios.
In light of these, a new \textit{scenario-adaptive} multiple-antenna NOMA paradigm can be formed towards NGMA.
It is worth noting that the proposed cluster-free NOMA framework provides a further generalization of the framework in our previous survey \cite{NGMA2022} 
and enables a higher SIC flexibility. 
Moreover, promising solutions for NGMA enabled by artificial intelligence (AI) will be discussed in this article.

\subsection{AI Enabled NOMA Towards NGMA}
Despite the promise of NGMA, the complicated multi-domain multiplexing also makes the interference suppression and the system optimization increasingly challenging.
Particularly, the communication design of next generation NOMA systems typically leads to a highly complex non-convex mixed-integer nonlinear programming (MINLP) problem, 
whose globally optimal solution is extremely difficult to obtain.
While conventional convex optimization methods can achieve the local optimum, they usually encounter several critical challenges in practice:
(i) They rely on sophisticated mathematical transformations and expert experiences to transfer original non-convex problems into tractable convex problems.
(ii) The resulting performance is severely sensitive to initialized parameters, which should be appropriately configured for different scenarios via laborious hand-engineered designs.
(iii) They often require large amounts of iterations to reach convergence, leading to an impractically high computational complexity especially in the overloaded regime 
or multi-cell networks.

Fortunately, recent advances in AI have created new opportunities to circumvent the above challenges, 
which enables the automated communication designs to combat the overwhelming system complexity and heavy dependence on human interventions 
\cite{AI_6G,L2O_2018,DeepUnfolding_2021,GNN_2021}.
In this article, we explore promising and advanced machine learning (ML) methods to empower NGMA via AI.
While existing research contributions have laid a solid foundation on multiple-antenna NOMA communication designs, AI enabled NGMA is still in its infancy, 
which motivates us to develop this treatise.
The main contributions of this article can be summarized as follows.
\begin{itemize}
  \item We propose a novel downlink cluster-free NOMA framework to achieve ultra-flexible SIC operations, which can generalize the existing multiple-antenna NOMA schemes, 
  thus enabling a new \textit{scenario-adaptive} communication paradigm towards NGMA.
  \item We investigate promising machine learning methods and highlight their features and application scopes, 
  which can autonomously learn high-efficiency communication designs with computationally realizable complexity and fewer human interventions.
  \item We propose two machine learning paradigms, which enables efficient SIC and beamforming design in single-cell and multi-cell networks 
  by extending conventional deep reinforcement learning (DRL) and graph neural network (GNN), respectively.
\end{itemize}

\section{AI Enabled NOMA Framework for NGMA}
In this section, we first propose a novel generalized cluster-free multiple-antenna NOMA framework to achieve \textit{scenario-adaptive} communications towards NGMA.
Then, to cope with the overwhelming complexity of next generation wireless systems, we further investigate promising AI enabled methods to achieve automated, 
computationally realizable, and efficient communication design.

\subsection{Cluster-Free NOMA Framework for NGMA}
As shown in Fig. \ref{fig_sys_model}, we propose a novel downlink cluster-free NOMA framework, which can be employed in both the single-cell and multi-cell networks.
Without loss of generality, we suppose that the base station (BS) is equipped with multiple antennas to serve multiple single-antenna users.
Note that the proposed framework can be applied to both underloaded and overloaded regimes.
To achieve fully spatial multiplexing, each user in our proposed framework is served by a dedicated beamforming vector.
Different from conventional multiple-antenna NOMA schemes that require sequentially performing SIC within predefined user clusters, 
the proposed framework completely eliminates the restriction of clusters and allows flexible SIC operations between users.
Specifically, the control of cluster-free SIC operation can be determined by a binary indicator, 
which decides whether a user will decode the signal of another user  before decoding its own signal for interference elimination.
Note that to ensure successful SIC decoding, the SIC decoding condition should be guaranteed \cite{CFNOMA_2022}.

The benefits of the proposed cluster-free NOMA are two-fold.
\textbf{(i) Generalization:} The proposed cluster-free NOMA provides a general framework.
Specifically, it is equivalent to SDMA/BB-NOMA when SIC operations are prevented/performed among all users.
Moreover, it reduces to CB-NOMA when SIC can be only sequentially carried out within predefined user clusters.
Therefore, the current SDMA, BB-NOMA, and CB-NOMA schemes that are suitable for specific scenarios can be considered as special cases of the proposed framework.
\textbf{(ii) Ultra-flexible SIC:} The proposed cluster-free NOMA provides an ultra-flexible SIC design.
It can adaptively prevent inefficient SIC among any users with weak channel correlations, 
thus avoiding system performance degradation caused by unfavourable SIC decoding conditions.
Meanwhile, it allows SIC between any highly channel correlated users to achieve flexible interference elimination without cluster limitations.
As a result, the proposed framework can enhance system performance and enable a novel \textit{scenario-adaptive} multiple-antenna NOMA paradigm towards NGMA.

\subsection{Promising Machine Learning Solutions for NGMA}

\begin{figure*}[!hbt]
  \vspace{-1.5em}
  \centering
  \includegraphics[width=0.85\textwidth]{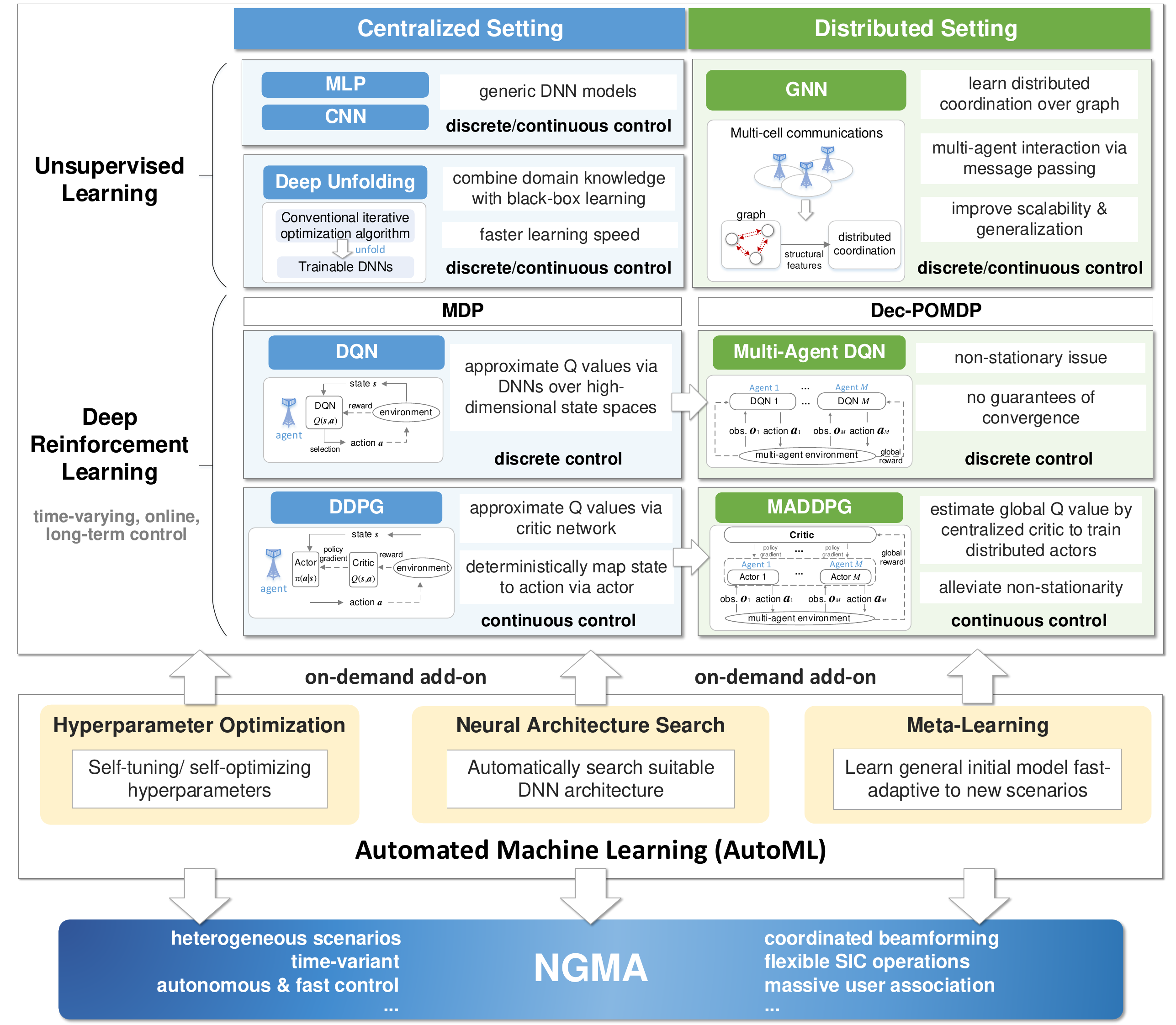}\\
  \caption{Overview of AI enabled NGMA.}\label{fig_L2O}
\end{figure*}

Due to the increasing complexity of next generation wireless systems, 
the communication design typically needs to solve challenging high-complexity, coupled, and non-convex MINLP problems.
To circumvent the inefficiencies of conventional convex optimization methods for NGMA, we investigate promising machine learning solutions in this part.

Relying on the outstanding ability of deep models to fit any arbitrary function, 
AI can train deep neural networks (DNNs) to approximate the optimal solutions of challenging optimization  problems. 
Specifically, the optimal solutions can be regarded as a non-convex function that maps the system state to the optimization variables.
AI can automatically learn high-quality solutions for communication designs in a data-driven manner, 
which avoids the dependence on expert knowledge and the hand-engineered parameter initialization required by conventional optimization methods.
Moreover, since DNNs only require lightweight computations during the forward propagation, 
learning-based solutions can significantly reduce the computational complexity to achieve fast time response and automated control in realistic environments.

Since supervised learning requires massive data samples labelled with high-quality solutions, it has limited applicability for complex optimization problems.
Hence, we focus on prospective unsupervised learning and DRL based solutions, and also investigate the emerging automated machine learning (AutoML) methods.
An overview of AI enabled NGMA can be illustrated in Fig. \ref{fig_L2O}, and the comparisons of representative machine learning algorithms are summarized in Table \ref{Table_alg}.

\begin{table*}[!htb]
  \begin{center}
  \caption{Comparisons of representative machine learning algorithms}\label{Table_alg}
  \label{tab1}
  \resizebox{0.8\linewidth}{!}{
  \begin{tabular}{ c  c  c  c  c c}
  \hline
  Algorithm & Time-varying             & Control         & Setting                   & Learning                                   & Data\\
  \hline
  MLP/CNN         & \XSolid                  & Both         & Centralized                        & Data-driven       & Non-structural \\
  deep unfolding  & \XSolid                  & Both         & Centralized                        & Model-based     & Non-structural \\
  GNN             & \XSolid                  & Both         & Centralized training, distributed execution with message passing  & Data-driven       & Structural \\
  DQN             & \Checkmark               & Discrete     & Centralized                        & Data-driven       & Non-structural \\
  DDPG            & \Checkmark               & Continuous   & Centralized                        & Data-driven       & Non-structural \\
  Multi-agent DQN    & \Checkmark            & Discrete     & Distributed training \& execution             & Data-driven        & Non-structural \\
  MADDPG          & \Checkmark               & Continuous   & Centralized training, distributed execution              & Data-driven  & non-structural \\
  \hline
  \end{tabular}}
  \end{center}
  \end{table*}

\subsubsection{Unsupervised learning}
While supervised learning aims to approximate the pre-labelled solutions, unsupervised learning can train the model parameters by directly minimizing unsupervised loss function without knowing ground-truth labels. 
Two generic machine learning models commonly employed in wireless communications are multi-layer perceptron (MLP) and convolutional neural network (CNN) \cite{L2O_2018}. 
However, since these models are inherently black-box and originally developed to accomplish computer vision tasks, they generally lead to degraded performance when being employed in wireless communications. 
To tailor unsupervised learning for wireless communications and circumvent these drawbacks, the following two approaches have been proposed.  
\begin{itemize}
  \item \textbf{Deep unfolding:} Deep unfolding provides a new learning paradigm to incorporate domain knowledge and optimization theory into learning \cite{DeepUnfolding_2021}. 
  The core idea is to unfold the iterative optimization algorithm as the layer-wise learning model. By approximating the iterative pattern of optimization, 
  black-box DNNs can be transformed into trainable white-box models with both theoretically interpretability and learning capability. 
  Hence, enhanced performance can be achieved while reducing the number of training parameters and accelerating the convergence.
  \item \textbf{Graph neural networks (GNNs):} Compared to conventional non-structural DNNs, 
  GNNs propose a structural learning paradigm to model the interplay effects and unveil the underlying dependencies among wireless nodes \cite{GNN_2021}. 
  Based on GNNs, the structural features over graph can be learned via message passing to enable distributed inference. 
  Therefore, GNNs can significantly improve scalability and generalization ability of unsupervised learning, thus realizing efficient coordination in distributed wireless communication designs.
\end{itemize}

\subsubsection{Deep reinforcement learning}
Reinforcement learning is typically applied for the long-term optimization problem that can be modelled into Markov decision process (MDP).
Specifically, each BS can be considered as an autonomous agent that interacts with environment to consistently improve its decisions through trial and error.
By observing the system state at each time slot, it decides the optimization variables, namely action, to maximize the accumulated discount reward.
DRL integrates deep learning with reinforcement learning, which utilizes DNNs for function approximations over high-dimensional state spaces.
Specifically, deep Q-network (DQN) \cite{DQN_2015} learns to estimate the Q value based on the Bellman equation, according to which the action can be greedily chosen.
Since DQN is only efficient for discrete control, 
deep deterministic policy gradient (DDPG) has been further proposed for continuous control based on the actor-critic framework \cite{DDPG_2015}.
DDPG trains an actor to deterministically map the states to continuous actions, whilst learning a critic to estimate the Q value function. 
For multi-cell distributed communication designs that can be modelled into decentralized partially observable MDP (Dec-POMDP), DQN can be directly extended to a multi-agent variant, 
where each agent independently performs Q-learning using local observations.
However, this would lead to a non-stationary environment for each agent, which violates the Markov assumptions and fails to guarantee the convergence of Q-learning.
Therefore, multi-agent deep deterministic policy gradient (MADDPG) \cite{MADDPG} has been further proposed. Based on centralized training and distributed execution, 
MADDPG trains multiple distributed actors with a centralized critic to achieve robust coordination.

Owing to the discrepancy between simulation and reality, directly applying the DRL policy trained from a simulated environment to the realistic environment may be inefficient. 
A typical technique to address this issue is domain randomization, 
where the DRL policies are trained with randomized distribution of perturbed simulator parameters to characterize the realistic distribution as much as possible. 
Moreover, to directly achieve the simulation-to-reality policy transfer, domain adaptation has been further proposed by invoking transfer learning. 
Specifically, the simulation and the reality are regarded as the source domain and the target domain, respectively.
By leveraging the structure similarity across different domains, the features, skills, and/or knowledge learned from the source domain can be transferred 
and adapted into the target domain with only a few number of realistic data samples.

\subsubsection{Automated machine learning}
Since next generation wireless systems are highly time-variant and comprise extremely heterogeneous applications, 
the parameters and the neural network architecture of the learning model should be both self-configured and fast-adaptive to various scenarios.
To automatically configure the learning models, the emerging paradigm of AutoML \cite{AutoML_2019} should be combined to enhance the machine learning based communication design, 
thus significantly reducing the human interventions and improving performance.
Overall, AutoML involves three techniques, namely hyperparameter optimization (HO), neural architecture search (NAS), and meta-learning \cite{AutoML_2019}.
On the one hand, HO and NAS can automatically optimize the hyperparameters and neural architecture of learning models, respectively.
On the other hand, meta-learning, also known as \textit{learning to learn}, aims to learn a general initial model that can be fast-adaptive to newly unseen communication scenarios.
According to the requirements of different application scenarios, these AutoML techniques can be adopted on demand as add-on modules to assist communications for NGMA.

\section{Efficient Machine Learning Paradigms for Cluster-Free SIC and Beamforming Design}
In this section, we explore novel machine learning paradigms for efficient cluster-free SIC and beamforming design based on the proposed cluster-free framework.
For both single-cell and the multi-cell networks, efficient centralized DRL and distributed GNN methods are proposed, respectively.

\subsection{Single-Cell Network: Centralized Learning Design}

We first investigate the centralized DRL solution to design the cluster-free SIC and beamforming in the single-cell network.
DRL can tackle the sequential MDP problem, where some kind of long-term average quality-of-service (QoS) metrics, such as time-average system sum rate, traffic delay, or outage probability, can be optimized in the time-varying online environment.
Specifically, the autonomous agent, i.e., BS, determines its action according to the DRL policy by observing sate at each time slot.
The state includes the information from the surrounding environment, e.g., channel state information (CSI) and users' newly traffics arrival at each time slot.
Moreover, the action includes both the combinatorial cluster-free SIC operations and continuous beamforming vectors, which leads to a discrete-continuous hybrid action space.

The main challenge in applying DRL for cluster-free NOMA design is that the native DRL methods cannot efficiently handle the high-dimensional hybrid action space, 
as analyzed as follows.
\textbf{(i) Hybrid:} Conventional DRL algorithms usually transform the hybrid action space into a unified space via discretization or relaxation.
The former discretizes continuous actions and learns over the discrete space, while the latter relaxing discrete actions and learning over the continuous space. 
However, both these techniques rely on approximations, which have limited performance.
\textbf{(ii) High-dimensionality:} Conventional DRL algorithms directly deal with the original high-dimensional combinatorial action space, 
which ignore the underlying structure and inherent dependencies of hybrid actions. 
This results in a poor scalability and dimension disaster in large-scale antenna systems or overloaded regimes.

\begin{figure}[!htb]
  \centering
  \includegraphics[width=0.5\textwidth]{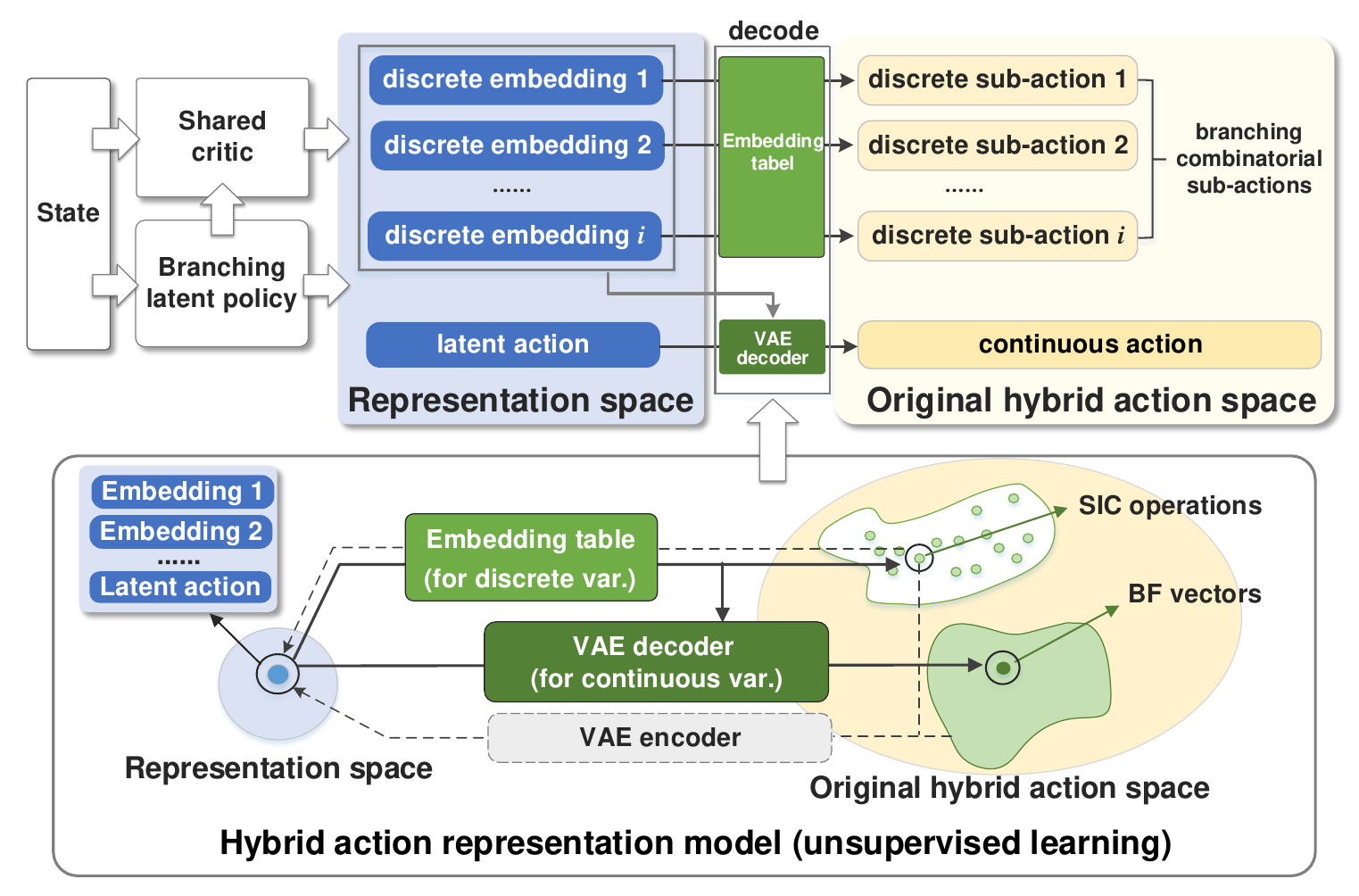}\\
  \caption{Illustration of the centralized BHy-DRL for single-cell cluster-free SIC and beamforming design.}\label{fig_BDRL}
\end{figure}

To address the weaknesses of native DRL methods,
we conceive a novel branching hybrid-action deep reinforcement learning (BHy-DRL) method based on the actor-critic framework. 
The diagram of BHy-DRL is shown in Fig. \ref{fig_BDRL}.
The key idea is to branch the high-dimensional combinatorial action into multiple low-dimensional sub-actions, 
and then invoke representation learning to capture the dependencies among the decomposed components of the hybrid action. 
To be more specific, BHy-DRL first decomposes the combinatorial action into $I$ sub-actions, where each sub-action $i$ includes the SIC operation between two users. 
Then, to capture internal dependencies between hybrid action components, 
both discrete and continuous components from original hybrid action space will be jointly encoded into a \textit{latent action} on the low-dimensional representation space. 
Different from conventional DRL that trains policies over the original hybrid action space, 
BHy-DRL learns a latent policy that outputs the \textit{latent action} over the compact representation space. 
To interact with the environment, this \textit{latent action} would be decoded back into the original hybrid action. 
We invoke unsupervised learning to jointly train the decoders and encoders.
For discrete components, an embedding table is constructed to encode each sub-action into an embedding. 
Then, for continuous components, a conditional Variational Auto-Encoder (VAE) is trained, 
which takes the system state and the embeddings as conditions to encode/decode the original/latent action, 
thus capturing dependencies between continuous and discrete actions.

\subsection{Multi-Cell Network: Distributed Learning Design}

The multi-cell network should mitigate both intra-cell and inter-cell interference.
When directly applying conventional algorithms, BSs should exchange their local CSI information, which may incur overwhelming communication burdens and data privacy concerns.
Furthermore, solving the resulting high-dimensional optimization problem usually leads to an intolerably high complexity.
To overcome these issues, we extend GNN to achieve efficient distributed design, where BSs can coordinate by exchanging low-dimensional messages that embeds their local information.
We propose a novel automated-learning GNN (AutoGNN) architecture, which can automatically learn efficient GNN architecture design to relieve the communication and computation overheads for coordination.

\begin{figure}[!htb]
  \vspace{-2em}
  \centering
  \includegraphics[width=3.5in]{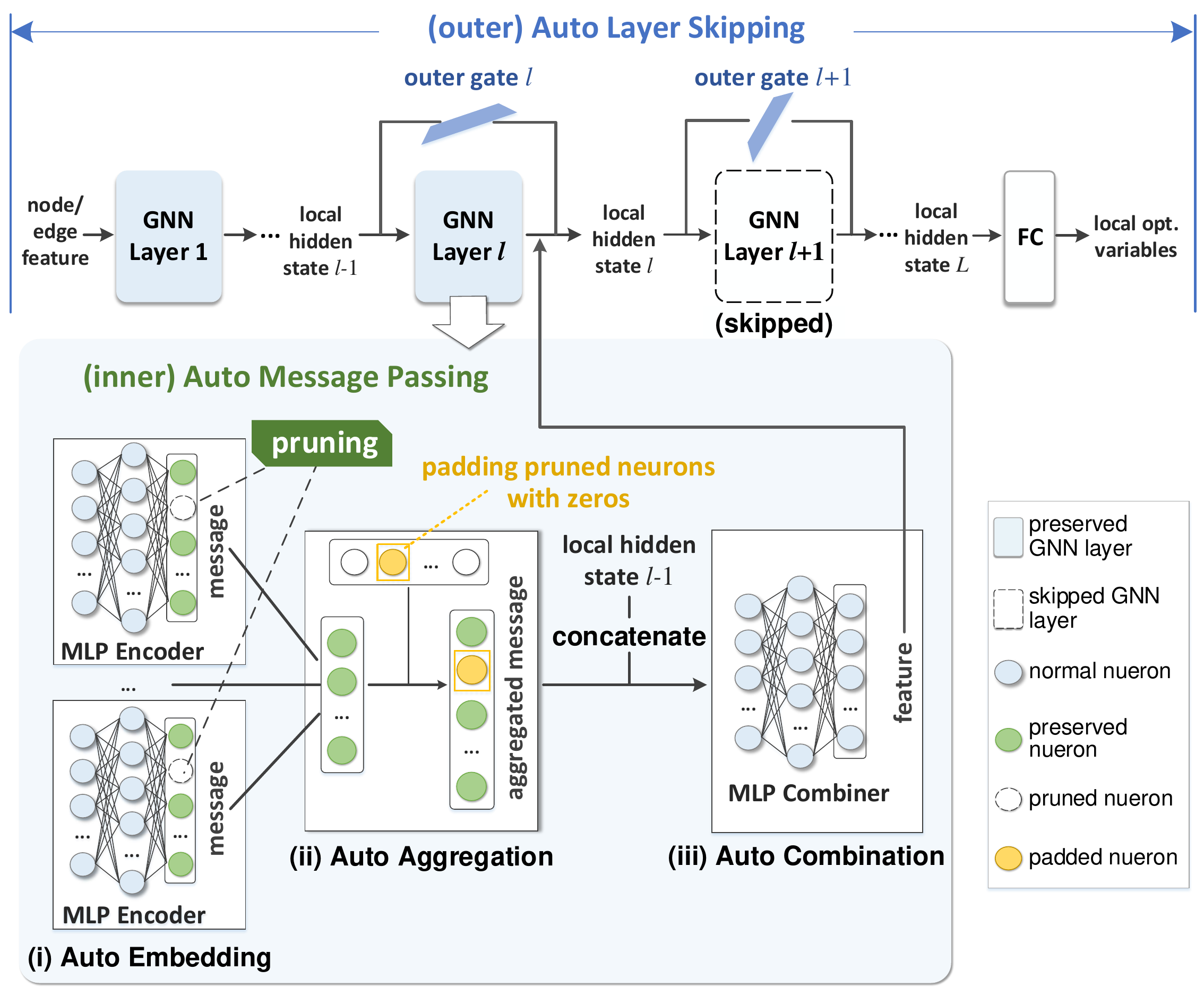}\\
  \caption{Illustration of the proposed distributed AutoGNN for multi-cell cluster-free SIC and beamforming design.}\label{fig_AutoGNN}
\end{figure}

We first review the conventional GNN architecture.
Specifically, the multi-cell environment is modelled by a graph, where graph nodes and edges represent BSs and their interactions, respectively.
The data channels and interference channels locally observed by each BS are respectively described as the node and edge features.
BSs can share the same GNN model parameters to reduce complexity.
Similar with MLPs and DNNs, GNN adopts a multi-layer structure.
Each GNN layer includes a communication round to implement message passing between agents, which has three stages as follows.

\textbf{(i) Message embedding:}  Each BS embeds local node/edge features into messages via an encoder implemented as MLP, and then passes the embedded messages to  neighbouring BSs. 
Here, the network width of the final MLP layer determines the size of the embedded message, which is termed \textit{message embedding size} in this paper.

\textbf{(ii) Message aggregation:} Each BS aggregates the messages received from neighbours via a permutation-invariant aggregation function, such as $\max()$, $\mathrm{sum}()$, and $\mathrm{mean}()$.

\textbf{(iii) Message combination:} Each BS updates local hidden state by combining the aggregated message with current hidden state and the node feature.
Eventually, it obtains local optimization variables through a fully connected (FC) layer.

\begin{figure}[!htb]
\vspace{-2em}
\centering
\subfloat[]{\includegraphics[width=3in]{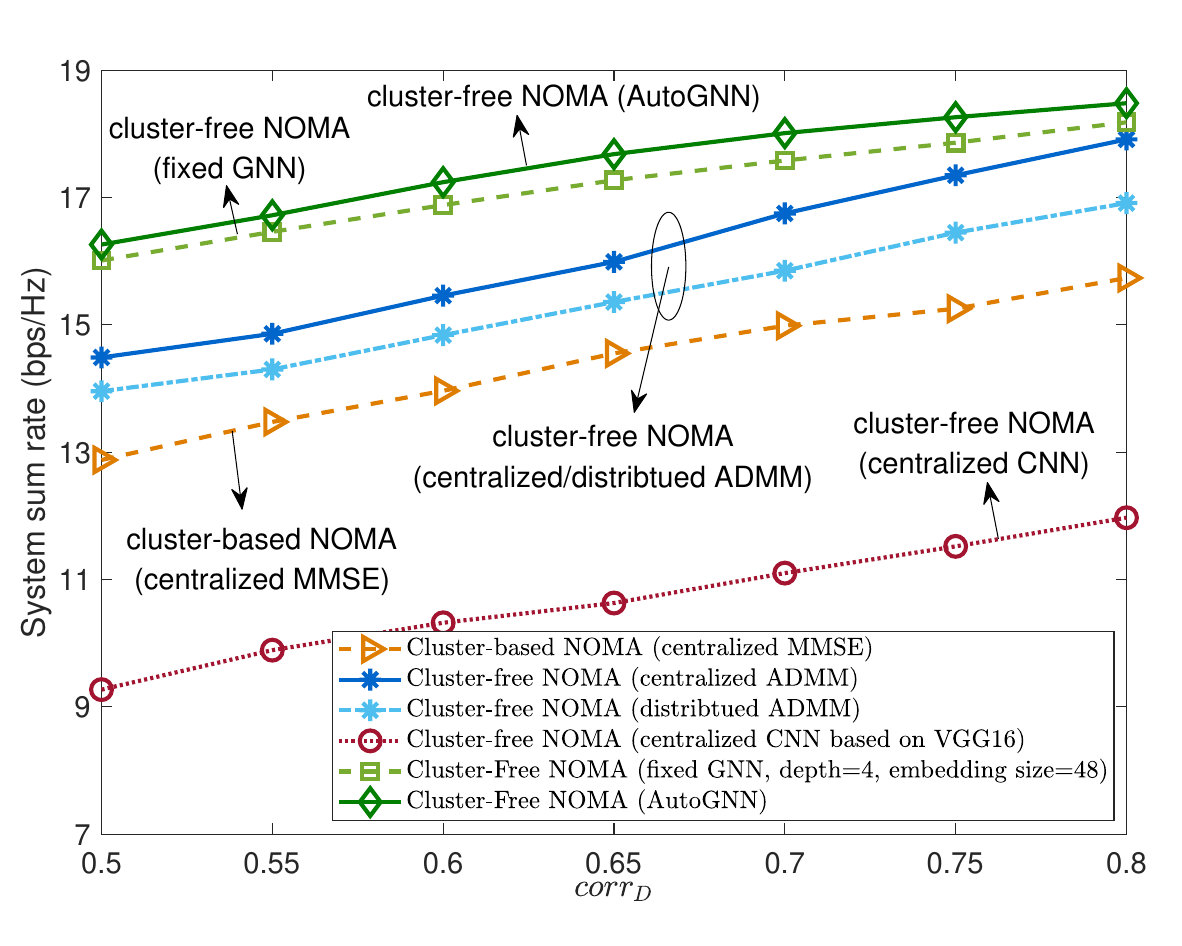}%
\label{fig_R_corr}}
\\ 
\vspace{-1em}
\subfloat[]{\includegraphics[width=3in]{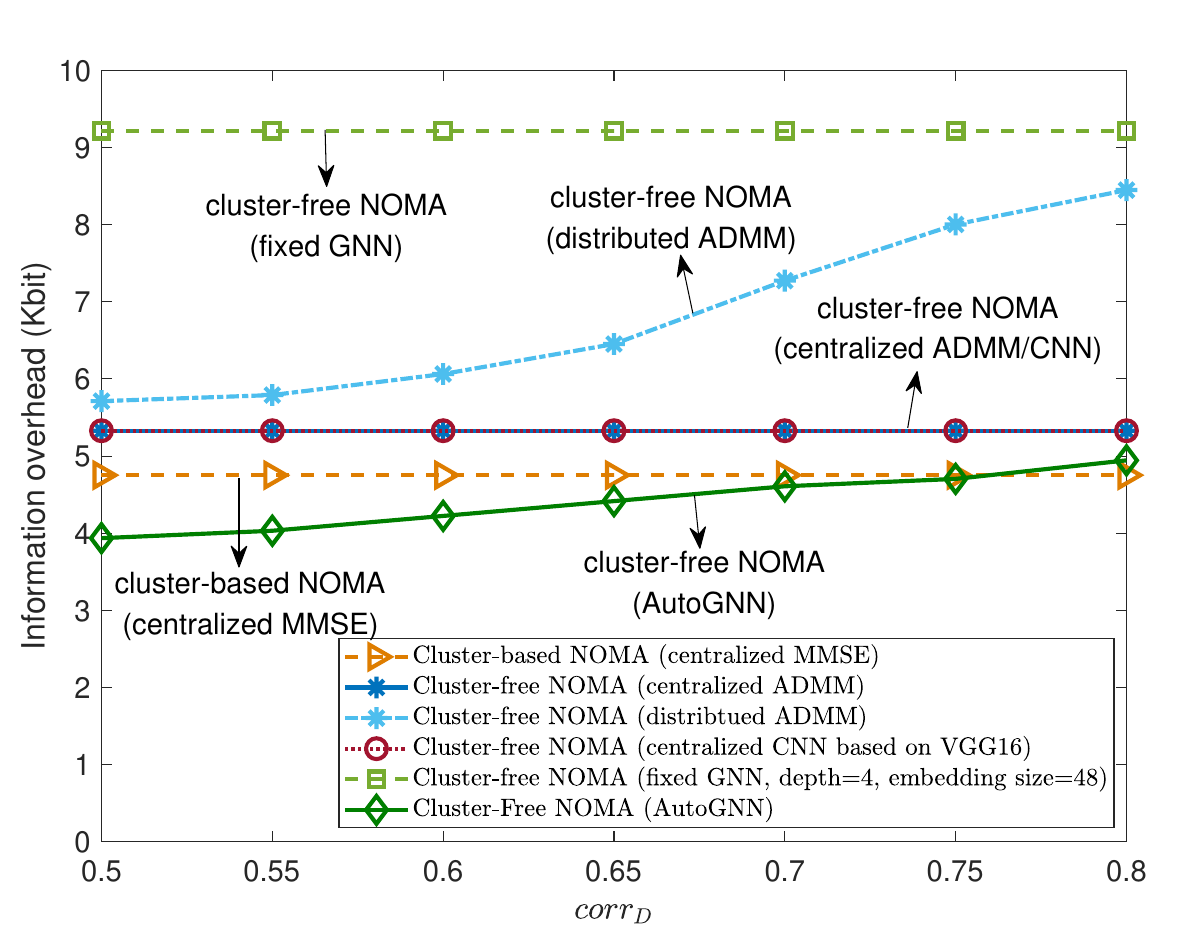}%
\label{fig_comm_cost}}
\caption{Performance comparisons of multi-cell multiple-antenna NOMA communications. (a) System sum rate versus data channel correlations $corr_{D}$. 
(b) Information overhead among BSs versus data channel correlations $corr_{D}$, where each real floating number takes up $8$ bits and each complex floating number occupies $16$ bits.}
\label{fig_comparison_AutoGNN}
\end{figure}

Despite enabling distributed coordination, GNN still inherits the black-box nature of deep learning.
Therefore, the neural network architecture is often cumbersome and awkward, which requires elaborate manual designs, intricate experiences and expertise.
Particularly, large GNN \textit{network depth} (namely the number of GNN layers) and \textit{message embedding size} will incur high communication and computation overheads.
However, if they are too small, the representation capability of the learning model may be insufficient, thus deteriorating the system performances.
To intelligently explore desirable GNN architecture, we propose a novel AutoGNN architecture for distributed multi-cell SIC and beamforming optimization, 
as shown in Fig. \ref{fig_AutoGNN}.
The proposed AutoGNN can automatically self-optimize the GNN architecture whilst training the GNN model parameters.
Inspired by NAS \cite{AutoML_2019}, we achieve the self-optimization via an auto-learned module, which has a dual-loop structure.
The GNN \textit{network depth} is designed in the outer loop, while the \textit{message embedding size} of each GNN layer is configured in the inner loop.
The auto-learned module and the GNN model parameters can be jointly learned based on the bi-level differentiable learning \cite{DARTS}.

Fig. \ref{fig_comparison_AutoGNN} verifies the performance of the proposed AutoGNN.
We consider $3$ BSs here. Each BS equips $4$ antennas to serve $6$ single-antenna users. 
Experimental details are omitted here due to the space limitation. 
The implementation code can be available at \href{https://github.com/xiaoxiaxusummer/AI_NOMA}{https://github.com/xiaoxiaxusummer/AI\_NOMA}.
As demonstrated in Fig. \ref{fig_comparison_AutoGNN}(a), 
the proposed cluster-free NOMA achieves higher performance than CB-NOMA under different channel correlations. 
For the joint optimization, the unsupervised centralized CNN yields the worst performance since it is non-structural and lacks scalability. 
In comparison, the structural GNN achieves a comparable system sum rate with conventional centralized optimization methods, i.e., 
centralized alternating direction method of multipliers (ADMM), whilst outperforming the distributed ADMM. 
As presented in Fig. \ref{fig_comparison_AutoGNN}(b), distributed ADMM can reduce the computational complexity and preserve data privacy, 
but incurs overwhelming information overheads since it requires exchanging intermediate information during each iteration. 
In contrast, centralized ADMM yields lower information overheads but the highest computational complexity. 
Compared to the fixed GNN and the centralized CNN, 
the proposed AutoGNN adaptively reduces communication burdens in various scenarios without degrading the system performance, which verifies its efficiency.

\section{Interplay Between Cluster-Free NOMA With Emerging Techniques}
In this section, we discuss the interplays between  cluster-free NOMA and pivotal next generation wireless techniques.
\subsection{Reconfigurable Intelligent Surface}
As a disruptive technology beyond 5G, reconfigurable intelligent surface (RIS) \cite{RIS_2020} can dynamically reconfigure the wireless environment and enable smart radios.
The interplay between RIS and cluster-free NOMA is analyzed as follows.
\begin{itemize}
  \item \textbf{RISs for cluster-free NOMA systems:} While cluster-free NOMA enabling higher flexibility, the system design is still subject to uncontrollable channel conditions.
RISs can improve the cluster-free NOMA systems from multiple aspects.
(i)  RISs can enhance network coverages and reduce outage probabilities of NOMA systems, especially in high-frequency bands.
(ii) RISs can provide extra spatial degree of freedom (DoF) to decrease users' channel correlations and mitigate interference, 
thus reducing the SIC decoding complexity as well as improving the spectral efficiency.
(iii) RISs can reconfigure the SIC implementations and the SIC decoding order by altering channel correlations and effective gains experienced by users, respectively. 
Thus, the cluster-free NOMA design can be tailored to users' QoS requirements, empowering high-quality user-centric communications.
  \item \textbf{Cluster-free NOMA for RISs-aided networks:} NOMA is indispensable to enhance user connectivity and realize massive access for RISs-aided networks.
      Since conventional multiple-antenna NOMA schemes are scenario-specific, the wireless environment should be accordingly reconfigured by RISs.
      In contrast, by enabling ultra-flexible SIC operations, 
      the generalized cluster-free NOMA can achieve adaptive interference management and efficient multi-domain multiplexing in various scenarios, 
      despite the correlations of the original/reconfigured spatial channels.
      Therefore, it can adaptively increase the spectral and energy efficiency of RISs-aided networks, especially under the non-ideal RIS operating conditions.
\end{itemize}

\subsection{MmWave/THz networks}
MmWave/THz networks can provide broadband spectrum resources from $30$ MHz to $10$ THz to enable multi-gigabit rates and low latency for next generation wireless systems.
However, in order to compensate the severe propagation loss over the high-frequency spectrum, 
mmWave/THz networks generally rely on the highly directional beamforming with the massive multiple-input multiple-output (MIMO) or extremely large-scale MIMO (XL-MIMO) techniques.
Due to the channel sparsity and the highly directional pencil-beam antennas, the mutual interference suffered by mmWave/THz users 
can be significantly reduced compared to sub-6GHz networks.
To guarantee successful SIC decoding without degrading system performances, the SIC operations should be only performed between users with sufficiently high channel correlations.
For this purpose, cluster-free NOMA can flexibly prevent detrimental SIC operations while only allowing beneficial SIC operations to promote mmWave/THz system performance.

It is worth mentioning that existing mmWave/THz massive NOMA schemes mostly rely on the far-field channel modelling. 
However, with XL-MIMO, the Rayleigh distance will increase with the number of antennas, 
thus forcing a shift from conventional plane-wave-based far-field communications to spherical-wave-based near-field communications.
Compared to far-field channels, near-field channels may experience energy spread effects, 
i.e., the energy of each path component may be dispersed into multiple angles and thus channel sparsity over the angular domain cannot be maintained.
This constitutes a research interest on cluster-free NOMA strategies in near-field MIMO channels, where the angular-domain sparsity no longer holds and the polar-domain sparsity should be considered.

\subsection{Integrated Sensing and Communication}
Next generation wireless systems are envisioned to provide both high-performance communications and high-precision sensing capabilities
for emerging wireless applications, such as smart cities and connected vehicles. 
This has given rise to the novel concept of integrated sensing and communication (ISAC), 
which supports both communication and sensing functions using a unified physical infrastructure and common spectrum resources. 
However, due to the fundamental trade-off between sensing qualities and communication performance, 
a crucial challenge for ISAC is to realize efficient resource sharing and interference coordination between communicating and sensing signals.
As an essential multi-domain multiplexing technology, multiple-antenna NOMA has become a key enabler for ISAC developments \cite{NOMAISAC}.
As a further advance, cluster-free NOMA can deliver ultra-flexible SIC operations between sensing users and communication users in various scenarios.
With the aid of machine learning, scenarios-adaptive inter-functionality coexistence can be achieved for ISAC.

\section{Concluding Remarks and Future Directions}
This article has proposed an AI enabled cluster-free NOMA framework to generalize existing multiple-antenna NOMA schemes, 
which enables ultra-flexible SIC operations and \textit{scenario-adaptive} NOMA communications towards NGMA.
Prospective machine learning solutions have been discussed, and their features and application scenarios were highlighted.
Furthermore, efficient centralized and distributed learning paradigms were proposed for the single-cell and multi-cell SIC and beamforming designs, respectively.
However, investigations on AI enabled NGMA are still at a very early stage.
Some existing open research issues can be exemplified as follows:

\begin{itemize}
  \item \textbf{Model-based constrained ML for NGMA:}
  The communication design for NGMA usually involves non-convex, coupled, and mixed-integer constraints. 
  Most machine learning algorithms penalize constraint violations into loss functions, or employ projection operations to obtain feasible solutions, 
  which have limited capabilities to strictly guarantee these constraints.
  Recently, Lagrangian dual method and interior point method have been introduced to achieve the model-based machine learning, 
  which demonstrated the potentials of guiding machine learning with constrained optimization theory.
  This inspires the research interests on model-based constrained machine learning for communication design of NGMA.
  \item \textbf{ML empowered dynamic multi-objective optimization for NGMA}:
  Due to the time-variant and heterogeneous natures of next generation wireless systems, 
  the communication design will encounter multiple conflicting optimization objectives or constraints regarding the system rate, energy consumption, traffic latency, outage probability, and so on.
  Moreover, when wireless environments change, these conflicting objectives and constraints may change over time, which poses a challenge to predict the changing Pareto optimal front.
  To this end, efficient multi-task machine learning methods should be investigated to empower dynamic multi-objective optimization.
  \item \textbf{Accelerating AutoML for NGMA:}
  While machine learning can predict desirable solutions through low-complexity forward propagation, 
  the model training via back-propagation generally requires massive data samples and incurs heavy computation burdens.
  Particularly, when using the AutoML techniques (such as meta-learning and NAS), the training process would become more time-consuming and computationally intensive.
  To relieve training costs, how to build high-performance lightweight models and accelerate AutoML remains a crucial but challenging research problem for NGMA.
 \end{itemize}

\end{document}